\documentclass[]{emulateapj}
\usepackage{natbib}
\usepackage{amsmath}
\usepackage{xspace}

\newcommand{\ergs}{ergs$^{-1}$\xspace}
\newcommand{\ml}{$M/L$\xspace}
\newcommand{\chandra}{{\it Chandra}\xspace}
\newcommand{\hst}{{\it HST}\xspace}
\def\Msun{\hbox{$\thinspace M_{\odot}$}\xspace}

\def\lk{\hbox{$\thinspace L_{\rm K}$}\xspace}
\def\lx{\hbox{$\thinspace L_{x}$}\xspace}

\shorttitle{Testing for variability in the IMF at high stellar masses}
\shortauthors{Peacock et al.}

\begin{document}

\title{Evidence for a constant IMF in early-type galaxies based on their X-ray binary populations$^{\dag \ddag}$}

\author{Mark B. Peacock$^{1}$, Stephen E. Zepf$^{1}$, Thomas J. Maccarone$^{2}$, Arunav Kundu$^{3,4}$, Anthony H. Gonzalez$^{5}$, \\Bret D. Lehmer$^{6,7}$, Claudia Maraston$^{8,9}$}
\affil{$^{1}$Department of Physics and Astronomy, Michigan State University, East Lansing, MI 48824, USA; mpeacock@msu.edu}
\affil{$^{2}$Texas Tech University, Physics Department, Box 41051, Lubbock, TX 79409, USA}
\affil{$^{3}$Eureka Scientific, Inc., 2452 Delmer Street, Suite 100 Oakland, CA 94602, USA}
\affil{$^{4}$Tata Institute of Fundamental Research, Homi Bhabha Rd, Mumbai 400005, India}
\affil{$^{5}$Department of Astronomy, University of Florida, Gainesville, FL 32611, USA}
\affil{$^{6}$The Johns Hopkins University, Homewood Campus, Baltimore, MD 21218, USA}
\affil{$^{7}$NASA Goddard Space Flight Center, Code 662, Greenbelt, MD 20771, USA}
\affil{$^{8}$Institute of Cosmology and Gravitation, Dennis Sciama Building, Burnaby Road, Portsmouth PO1 3FX, UK}
\affil{$^{9}$South East Physics Network, www.sepnet.ac.uk}

\email{$^{\dag}$Based in part on observations made with the NASA/ESA {\it Hubble Space Telescope}, and obtained from the Hubble Legacy Archive, which is a collaboration between the Space Telescope Science Institute (STScI/NASA), the Space Telescope European Coordinating Facility (ST-ECF/ESA) and the Canadian Astronomy Data Centre (CADC/NRC/CSA).\\
$^{\ddag}$The scientific results reported in this article are based in part on data obtained from the \chandra Data Archive and observations made by the \chandra X-ray Observatory and published previously in cited articles.}

\begin{abstract}
\label{sec:abstract}

A number of recent studies have proposed that the stellar initial mass function (IMF) of early type galaxies varies systematically as a function of galaxy mass, with higher mass galaxies having bottom heavy IMFs. These bottom heavy IMFs have more low-mass stars relative to the number of high mass stars, and therefore naturally result in proportionally fewer neutron stars and black holes. In this paper, we specifically predict the variation in the number of black holes and neutron stars based on the power-law IMF variation required to reproduce the observed mass-to-light ratio trends with galaxy mass. We then test whether such variations are observed by studying the field low-mass X-ray binary populations (LMXBs) of nearby early-type galaxies. In these binaries, a neutron star or black hole accretes matter from a low-mass donor star. Their number is therefore expected to scale with the number of black holes and neutron stars present in a galaxy. We find that the number of LMXBs per K-band light is similar among the galaxies in our sample. These data therefore demonstrate the uniformity of the slope of the IMF from massive stars down to those now dominating the K-band light, and are consistent with an invariant IMF. Our results are inconsistent with an IMF which varies from a Kroupa/Chabrier like IMF for low mass galaxies to a steep power-law IMF (with slope $x$=2.8) for high mass galaxies. We discuss how these observations constrain the possible forms of the IMF variations and how future \chandra observations can enable sharper tests of the IMF.
%based on the IMF variation required to reproduce the observed mass-to-light ratio trends with galaxy mass. 

\end{abstract}


\section{Introduction}
\label{sec:intro}

The stellar initial mass function (IMF) describes the initial distribution of masses when a population of stars formed. The IMF is of fundamental importance to a wide range of astrophysics. Unfortunately, it is very difficult to directly measure the IMF for external galaxies, and our knowledge of the IMF is primarily based on the studies of the Milky Way (MW), where the stellar population can be directly measured to low stellar masses \citep[e.g.][]{Kroupa01,Chabrier03}. Written as a differential mass function, $dN/dm \propto m^{-x}$, the Kroupa Galactic IMF has a Salpeter-like slope of $x$=2.3 \citep{Salpeter55} for stars more massive than $0.5 \Msun$ and a flatter slope of $x$=1.3 for stars with masses between 0.5 $\Msun$ and 0.08 $\Msun$. The Chabrier log-normal representation of the IMF is very similar. Studies of the MW's IMF have shown it to be generally invariant \citep[see e.g.][]{Bastian10} and a universal IMF, based on the MW's stellar population, is commonly assumed. However, in extragalactic studies some recent evidence suggests that the IMF may not be universal.

One method for investigating the ratio of low to high mass stars in unresolved stellar populations is to look at the strength of gravity-sensitive features in their integrated spectra \citep[e.g.][]{Cohen78,Faber80,Carter86,Couture93}. Some such studies have found that the strengths of the giant sensitive Ca{\sc ii}~triplet indices decrease \citep[][]{Saglia02,Cenarro03} and the dwarf sensitive Na{\sc i}~doublet and Wing-Ford molecular FeH band absorption features increase \citep{vanDokkum10,vanDokkum11} with galaxy velocity dispersion. These observations, and full spectral fitting to stellar population models, have led to a number of papers suggesting that the IMF may become increasingly bottom heavy as the luminosity and velocity dispersion of the galaxy increases \citep{Conroy12b,Ferreras13,LaBarbera13}. Specifically, these papers find that lower mass elliptical galaxies have spectra consistent with a Kroupa like IMF (as seen in the MW), while galaxies at the high mass end require a steeper IMF, with slopes up to $x\simeq3$ proposed for a single power-law. More complex IMF models, such as a broken power-law, where only the low mass slope steepens \citep[][]{Conroy12b} or a time-dependent IMF \citep{Weidner13} have also been proposed to explain these observations. 

Systematic variations in the IMF have also been proposed to explain the observed ``fundamental plane'' of elliptical galaxies \citep{Dressler87, Djorgovski87}. It has been known since the discovery of the fundamental plane that the observed scaling between velocity dispersion and the parameters relating to the effective radius and surface brightness deviates from that expected from the virial theorem. This deviation is such that the \ml for early-type galaxies increases systematically with increasing galaxy mass, luminosity, and velocity dispersion. The reason for this relationship has been investigated by many studies, with two mechanisms generally proposed \citep[e.g.][]{Renzini93, Zepf96, Treu10, Graves10, Cappellari12}. The first, is that higher velocity dispersion galaxies could have systematically larger fractions of dark matter in the inner regions. The alternative, is that the IMF may vary, with higher luminosity and larger velocity dispersion galaxies having a systematically larger \ml ratio because of systematic changes in the IMF. \citet{Cappellari12} proposed, based on their detailed dynamical investigation, that the dark matter variations can not explain the observed \ml variations. Their conclusion is that variations in the IMF are required to explain the observed \ml variations, with high mass galaxies having either a relatively top heavy IMF (with a slope $x$=1.5, due to relatively more stellar remnants contributing to the mass of the galaxy, but little to its light) or a relatively bottom heavy IMF (with a slope $x$=2.8, due to relatively more low mass stars, which have higher \ml ratios). The kinematic results cannot distinguish between these two cases, but are consistent with the bottom heavy IMF proposed from the absorption line studies cited above. 

Further evidence for a bottom heavy IMF in massive galaxies comes from some gravitational lensing studies. \citet{Treu10} studied 56 gravitational lenses and found that their inferred masses relative to those predicted from stellar population fits increased as a function of the galaxy's velocity dispersion. They tentatively conclude that this may be due to a steepening of the IMF as a function of galaxy mass. However, not all gravitational lensing results agree with a bottom heavy IMF in high mass galaxies. In particular, \citet{Smith13} recently studied the closest known strong-lensing galaxy, the giant elliptical ESO325-G004. This galaxy has a high velocity dispersion and features a strong (dwarf star sensitive) Na~I~8200${\rm \AA}$ spectral feature. However, the inferred mass to light ratio for this massive galaxy is consistent with that predicted from stellar population models with a standard Kroupa IMF and inconsistent with a Salpeter (or steeper) IMF. 

For very low mass galaxies, direct observations of the stellar populations in two of the Milky Way's dwarf spheroidal satellite galaxies suggests that they have a flatter than Kroupa IMF \citep[with $x$=1.2 for Hercules and $x$=1.3 for Leo IV;][]{Geha13}. Given the very low mass of these galaxies, this result is consistent with the proposed flattening of the IMF with decreasing galaxy mass. A flat IMF may even extend to the very low mass ultracompact dwarf galaxies (UCDs). Some UCDs are observed to have high \ml ratios that could be explained by them having either a relatively flat IMF or a dark matter component \citep[e.g.][]{Hasegan05,Mieske08a,Mieske08b}. \citet{Dabringhausen12} also proposed that UCDs may host more low mass X-ray binaries (LMXBs) than expected. If this is the case, it would be consistent with the higher fraction of stellar remnants that are produced by a flatter IMF. 

\begin{table*}
 {\centering
  \caption{Galaxy sample \label{tab:galaxy_data}}
  \begin{tabular}{@{}lcccrccccccc@{}}
%   \hline
   \hline
   \vspace{-2mm}
  \\
Name & type$^{\rm i}$ & dist.$^{\rm ii}$ & ref$^{\rm ii}$ & $S_{N}^{\rm iii}$ & log$({\rm \sigma_{1kpc}})^{\rm iv}$ & ref$^{\rm iv}$ & $r_{inner}^{\rm v}$ & $r_{ext}^{\rm vi}$ & $e^{\rm vi}$ & ${\rm M_{K}}^{\rm vi}$ & (J-K)$^{\rm vi}$  \\
NGC...&        &  Mpc &   &   &  kms$^{-1}$          &      & arcsec & arcsec & & &  \\
   \hline
   4649  & E2  & 16.5  & 2 & 6.7 &  2.488& 1 & 15   & 241.3 & 0.19 & -25.26 & 0.939  \\
   4472  & E2  & 16.7  & 2 & 5.6 & 2.460 & 1 & 15   & 313.4 & 0.19 & -25.61 & 0.883 \\
   1399  & E1  &  20.0 & 1 &12.4 & 2.447& 2 & 10   & 202.2 & 0.00 & -25.19 & 0.924\\
   4594  & SA  &  9.0  & 3 & 2.0 &  2.400 & 3 &22.5*& 297.1 & 0.46 & -24.76 & 0.993\\
   4278  &E1-2& 16.1 & 1 & 6.9 & 2.358 & 1 & 10   & 155.0 & 0.07 & -23.76 & 0.915  \\
   3379  & E1  &  10.6 & 1 & 1.2 & 2.294 & 1 & 10   & 191.7 & 0.15 & -23.53 & 0.907 \\
   4697  & E6  & 11.7  & 1 & 2.5 &  2.256& 1 & 10   & 240.2 & 0.37 & -23.85 & 0.880  \\
   7457  &SA0 & 13.2  & 4 & 3.1 &  1.870& 1 &  5    & 155.1 & 0.45 & -22.43 & 0.890  \\
   \hline
   \end{tabular}\\
 }
\footnotesize
\vspace{1mm}
Properties of the galaxies studied in this paper, sorted by decreasing $\sigma$. \label{tab:galaxy_data}
$^{{\rm i}}$from \citet{deVaucouleurs91}; 
$^{{\rm ii}}$distances in Mpc from surface brightness fluctuation measurements from: 
(1) \citet{Blakeslee01}; (2) \citet{Blakeslee09}; (3) \citet{Jensen03}; (4) \citet{Tonry01}; 
$^{{\rm iii}}$Globular cluster specific frequency from \citet{Ashman98} and \citet[][for NGC~7457]{Hargis11};
$^{{\rm iv}}$galaxy's velocity dispersion from: (1) \citet{Cappellari12}; (2) \citet{Saglia00}; (3) \citet{Jardel11};  
$^{{\rm v}}$The radius defining the central region that is excluded from our analysis *For NGC~4594 we remove an elliptical inner region with semi-minor axis = 22.5\arcsec and semi-major axis = 168\arcsec; 
$^{{\rm vi}}$Galaxy data from the two micron all sky survey (2MASS) large galaxy atlas (LGA) \citep{Jarrett03}, `total' extrapolated galaxy semi-major axis (${\rm r_{ext}}$), K$_{\rm s}$-band ellipticity ($e=1-b/a$), K$_{\rm s}$-band magnitude within this ellipse (${\rm M_{K}}$) and J-K${\rm _{s}}$ color \\
\end{table*}
%other 1399 sigma ref \citet{Gebhardt07}

\begin{table*}
 {\centering
  \caption{Galaxy data and X-ray source populations\label{tab:galaxy_data2}}
  \begin{tabular}{@{}lcccccccccc@{}}
%   \hline
   \hline
   \vspace{-2mm}
  \\
         & \multicolumn{3}{c}{\hst ACS data$^{{\rm i}}$} & \multicolumn{3}{c}{\chandra data$^{{\rm ii}}$} & Galaxy light$^{{\rm iii}}$ & \multicolumn{3}{c}{Number of X-ray sources$^{{\rm iv}}$}\\
Name & blue filter & red filter & No.  &  exp. time & 90$\%$ limit & ref    & covered & ${\rm N_{x,field}}$ & ${\rm N_{x,GCs}}$ & ${\rm N_{x,back}}$ \\
NGC...&      &  &  &   ksec   & $\times$10$^{38}$ergs$^{-1}$    &  & ${\rm L_{K,cov}/L_{K,ext}}$ &  &  &  \\
   \hline
   4649  & F475W & F850LP & 6 & 300  &  0.60 & 6  & 0.61 & 162 & 132 & 17 \\
   4472  & F475W & F850LP & 3 & 380  & 0.90  & 4  & 0.54 &  74 &  60 &  16 \\
   1399  &     -     & F606W & 9 & 101  & 1.20  & 1  & 0.78 &  62 &  85 &  28 \\
   4594  & F435W & F625W  & 6 & 174  &  0.20 & 5  & 0.42 &  74 &  39 &  18 \\
   4278  & F475W & F850LP & 4 & 458  & 0.10  & 3  & 0.66 &  84 &  79 &  26 \\
   3379  & F475W & F850LP & 4 & 324  & 0.05  & 2  & 0.75 &  68 &  14 &  19 \\
   4697  & F475W & F850LP & 7 & 132  &  0.14 & 7  & 0.81 &  47  &  31 & 15 \\
   7457  & F450W*& F814W*& 2*&  30   & 0.90  & 8  & 0.90 &    3  &   0  &  0 \\

   \hline
   \end{tabular}\\
 }
\footnotesize
\vspace{1mm}
$^{{\rm i}}$The number of different \hst/ACS fields used to identify optical counterparts to X-ray sources (No.) and the filters used. *For NGC~7457, only WFPC2 observations were available, the blue and red filters listed cover different regions of the galaxy, so allow source identification, but not color information; 
$^{{\rm ii}}$\chandra data used to study the galaxy. Quoted are the total exposure times, estimated 90$\%$ completeness limits and references for the X-ray catalog used. X-ray source catalog references are: (1) \citet{Paolillo11}; (2) \citet{Brassington08}; (3) \citet{Brassington09}; (4) \citet{Joseph13}; (5) \citet{Li10}; (6) \citet{Luo13}; (7) \citet{Sivakoff08}; (8) \citet{Gultekin12}; 
$^{{\rm iii}}$The fraction of the galaxy's stellar light covered by this study (relative to ${\rm M_{K}}$, quoted in Table \ref{tab:galaxy_data}); 
$^{{\rm iv}}$The number of X-ray sources: with no optical counterparts (${\rm N_{x,field}}$); associated with globular cluster (GC) like counterparts (${\rm N_{x,GCs}}$); and associated with other optical counterparts (${\rm N_{x,back}}$).
\\
\end{table*}

In this paper, we search for an independent signature of a variation in the IMF with galaxy mass. In particular, we probe the high mass end of the stellar populations in these galaxies, based on their field LMXB populations. These binaries consist of a black hole (BH) or a neutron star (NS) accreting from a low mass donor star and hence track the population of massive stars that formed in a galaxy. The proposed variation of a galaxy's IMF from a Kroupa IMF at low mass to an $x$=2.8 IMF at high mass therefore predicts relatively fewer LMXBs per stellar mass in higher mass galaxies. In Section \ref{sec:sample} we discuss the galaxies studied in this paper and the archived optical and X-ray data used. In Section \ref{sec:results}, we present the population of field LMXBs in these galaxies. Finally, in Section \ref{sec:imf} we predict how the LMXB population should vary as a function of mass due to a variable IMF and test these predictions against the observed populations.

\section{Galaxy sample $\&$ data} 
\label{sec:sample}

To investigate the LMXB populations of local galaxies, we select a sample of galaxies based on the following criteria: (1) they are early type galaxies with little ongoing star formation and thought to have similarly old stellar populations \citep[see e.g.][and Section \ref{sec:other}]{Trager00, Terlevich02, Sanchez-Blazquez06b,Silchenko06,Thomas10}; (2) they have precise dynamical mass estimates from \citet{Cappellari12}, \citet[][NGC~4594]{Gebhardt07} or \citet[][NGC~1399]{Jardel11}; (3) they have deep X-ray observations from the \chandra observatory, so that their LMXB populations can be reliably measured; (4) they have optical photometry from {\it Hubble Space Telescope} (\hst) advanced camera for surveys (ACS) mosaics covering most of the galaxy's optical emission. We exclude M~87 from this sample due to concerns over accurately measuring its LMXB population against the high background from the hot gas that makes up its interstellar medium. The resulting sample of galaxies is shown in Table \ref{tab:galaxy_data}. It can be seen that the galaxies span only a small range of colors. The sample includes the brightest cluster galaxies, NGC~4472 and NGC~1399 (also the central dominant galaxy in the Fornax cluster). These brightest galaxies are where the extremely bottom heavy IMF's have been previously proposed \citep{vanDokkum10}. The galaxies then span a range of K-band luminosities (\lk), down to lower mass galaxies that are thought to have Kroupa like IMFs, with \lk varying from $2-40\times10^{10}L_{K,\odot}$. The sample therefore probes the range of masses over which significant variations in the IMF should be present (if such variations exist). 

\subsection{X-ray data/ catalogs}

The total \chandra exposure times and associated detection limits for these galaxies are quoted in Table \ref{tab:galaxy_data2}. Most of the galaxies have long combined exposures of over 100ks. We also include the galaxy NGC~7457, which has a shorter exposure time of 30ks. This galaxy is of particular interest to our study because its relatively low mass should result in a large effect on its LMXB population, if the IMF relationship is present. These data allow the X-ray populations of these galaxies to be studied down to detection limits of $L_{x} = 3\times10^{36} - 1\times10^{38}{\rm ergs^{-1}}$ (where these limits are the 90$\%$ completeness limits that were determined by the studies referenced in Table \ref{tab:galaxy_data2}). This is deep enough to allow accurate measurements of the galaxies LMXB populations. These detection limits are taken from the papers cited in Table \ref{tab:galaxy_data2} and are in reasonably good agreement with the limits predicted by the web based simulator {\sc pimms} (v4.6a)\footnote{http://cxc.harvard.edu/toolkit/pimms.jsp}. 

The X-ray data available for all of these galaxies have previously been analyzed and published (see references in Table \ref{tab:galaxy_data2}). We do not repeat the previous analysis of these data, but take the X-ray source catalogs (XSCs) for each galaxy from the literature. This includes the quoted source locations\footnote{For NGC~3379, we use higher accuracy RA coordinates than the rounded values quoted in the paper. We thank Nicola Brassington for providing us with this catalog.} and X-ray luminosities (\lx). For NGC~3379, NGC~4278, NGC~4472 and NGC~4649 we take \lx as quoted in the papers, for NGC~4594 and NGC~4697 we convert the source counts to \lx using the conversions quoted in the papers. For NGC~1399, we convert the quoted flux in photons/s to \lx by comparing with the catalog of \citet{Liu11}. \citet{Liu11} provide a catalog of X-ray point sources in 383 nearby galaxies, including all of those considered here. Investigation of this catalog showed that it is not as complete as those provided by the individual studies, so it is not used as the primary dataset for any of the galaxies in our study. However, it does provide a relatively homogeneous catalog with which to compare the fluxes of sources quoted in each galaxy's XSCs. This comparison confirms that there are no large systematic offsets between the \lx values quoted by \citet[][in the band 0.3-8 keV]{Liu11} and those found by the different studies for all galaxies except NGC~4472. For NGC~4472, we find that the luminosities from the catalog of \citet{Joseph13} are fainter than those of \citet{Liu11} and the previous study of \citet[][which was based on shallower data]{Maccarone03}. A potential reason for this is that \citet{Joseph13} list source fluxes over the narrower 0.5-5keV range. For this study, we use the deeper catalog of \citet{Joseph13}, but scale the quoted \lx by 1.4 to match those quoted by \citet{Maccarone03} and \citet{Liu11}. 

We restrict our analysis of each galaxy to inside the $r_{ext}$ ellipse (as defined in Table \ref{tab:galaxy_data}). This ellipse defines the distance to which the galaxy's light can be extrapolated in 2MASS observations of the galaxy \citep[taken from the 2MASS LGA,][]{Jarrett03}. Field LMXBs may reside at greater distances than this from the center of some of these galaxies. However, the ratio of field sources to background and GC sources becomes very low at these larger distances. We also restrict our study to regions outside the central regions (given by $r_{inner}$ in Table \ref{tab:galaxy_data}). Inside of this region source confusion and gas emission can effect the reliability and detection limits of the XSCs. It is also harder to reliably associate X-ray sources with optical counterparts in these inner regions (where the density of both is very high). For NGC~4594, we restrict our analysis to the region outside of an inner ellipse in which the dust lane makes association with optical counterparts less reliable. For NGC~4649, we remove an additional region which covers X-ray sources within the D25 ellipse of the nearby galaxy NGC~4647. This is the same region that was excluded from the galaxy's XSC by \citet{Luo13}. The field LMXB population in a galaxy has been observed to trace its stellar emission \citep[e.g.][]{Kundu07}, hence removing regions from these galaxies should not influence our results. The fraction of each galaxy's K-band light that is covered by our study (relative to the total within $r_{ext}$) is quoted in Table \ref{tab:galaxy_data2}. This was calculated directly from the 2MASS LGA images of each galaxy by masking out the regions inside $r_{inner}$, outside of $r_{ext}$, and those regions that are not covered by \hst observations (which are important for removal of non field LMXBs from the XSCs, as discussed below).

\subsection{Optical counterparts}
\label{sec:matching}

For all galaxies, the XSCs used include all sources detected. They therefore include not only the population of  LMXBs associated with the field of these galaxies (which is desired for this study), but also background AGN and LMXBs located in GCs. Because all of the galaxies have very low levels of star formation, the presence of high mass X-ray binaries in the galaxies should be negligible. It is well established that a large fraction (20-70$\%$) of the LMXBs in these galaxies are located in GCs \citep[e.g.][]{Angelini01,Kundu02,Jordan04}. These LMXBs are likely formed via dynamical interactions \citep[e.g.][]{Clark75, Jordan07, Peacock10b}. The formation of LMXBs through dynamical processes increases with the stellar density, $\rho^{2}$ \citep[e.g.][]{Fabian75}. Formation through these mechanisms is therefore dominant in the cores of GCs, which have extremely high stellar densities, but insignificant in the fields of these galaxies, where the stellar density is orders of magnitude lower\footnote{We note that there is tentative evidence for some dynamical formation of LMXBs in the very central region of M~31 \citep{Voss07}. However, this would make only a small contribution to the total LMXB population in the galaxy and, in this paper, we exclude these central regions from our analysis.} \citep[see e.g.][]{Fabian75,Verbunt87}. Thus the GC LMXB population represents a different origin from the field LMXBs. To obtain reliable field populations, it is therefore vitally important to remove the GC LMXBs from our analysis. 

To remove X-ray sources associated with GCs (and background galaxies), we restrict our study to regions of these galaxies that have \hst/ACS photometry, and remove sources with optical counterparts. At the distances of these galaxies, the \hst observations have z-band detection limits 3-4 magnitudes fainter than the peak of the GC luminosity function \citep[GCLF; which peaks in the z-band at around -8.5, e.g.][]{Villegas10}. Since the GCLF extends to about 3-4 magnitudes fainter than this peak, we detect all but the very faintest GCs. Furthermore, LMXBs are primarily found in brighter, more massive GCs \citep[e.g.][]{Kundu03,Kim06}. We therefore expect to detect the optical counterpart to all of the GC LMXBs. The data used for each galaxy are listed in Table \ref{tab:galaxy_data2}. For all galaxies except NGC~4472 and NGC~7457, \hst/ACS mosaics are available that cover the vast majority of the galaxy out to $r_{ext}$. For NGC~4472, we cover a smaller fraction of the galaxy light by using only the three ACS fields that are available for the galaxy. For NGC~7457 only two pointings, taken with the wide field planetary camera 2 (WFPC2), are available - although these cover all of the detected X-ray sources. 

The \hst/ACS images were taken from the HLA\footnote{http://hla.stsci.edu/}, if available, or the MAST\footnote{http://archive.stsci.edu/hst/} archive, otherwise. We use the pipeline reduced and drizzle combined products that are provided by these archives. Background light, associated with the field stars in the galaxies, was subtracted from these images using a ring median filter using the {\sc iraf} task {\sc rmedian} with an inner radius of 30~pixels. The world coordinate system (WCS) of the images was then aligned relative to the XSCs using the {\sc iraf} task {\sc tfinder}. This was done interactively under the task by matching X-ray sources to likely counterparts in the images. Because a large fraction of the sources are associated with GCs or background galaxies, enough sources with optical counterparts were present in each ACS field to align its WCS to that of the relevant XSC's. 

Sources were identified and measured in these background subtracted and WCS aligned ACS images using {\sc Sextractor} \citep{Bertin96}. This was run with a detection threshold of 3$\sigma$ and using the associated weight images (the `WHT' images produced by the pipeline) to estimate the noise for each pixel in the science image. For all galaxies except NGC~1399, a red and blue filter was available for each field. We use the redder band as our primary source catalog and match sources to the bluer band where detected. Photometry was obtained through a 0.25$\arcsec$ aperture and calibrated to the AB system using the standard calibration, as described in the ACS data handbook \citep{Gonzaga13}. Our primary interest in finding the colors of sources is to identify GCs in the galaxies. These clusters should be marginally resolved. We therefore produced an empirical aperture correction for the 0.25$\arcsec$ aperture using the ratio of fluxes of bright sources through a 0.25$\arcsec$ and a 0.5$\arcsec$ aperture. The fluxes were then further corrected for losses from 0.5$\arcsec$ aperture to infinity assuming the aperture losses of a point source, as quoted by \citet{Sirianni05}. Finally, the sources were derredened using the Galactic extinction maps of \citet{Schlegel98}. 

Within our final optical catalog, sources are flagged as GC candidates based on (1) having colors in the range $0.6<g\!-\!z<1.6$ ($0.6<B\!-\!r<1.4$ for NGC~4594); (2) having absolute magnitudes in the range $-12.5<z<-6.5$ ($-12.0<r<-6.5$ for NGC~4594); (3) being extended based on a {\sc Sextractor} stellarity flag $<$0.9 or a difference in magnitudes measured between the 0.25$\arcsec$ and 0.5$\arcsec$ apertures of $m_{0.25}\!-\!m_{0.5}> 0.4$; (4) being not too extended to be a GC, $m_{0.25}\!-\!m_{0.5}<1.0$. For NGC 1399, for which we have only one filter, GC candidates are selected based on criteria 2-4 only. In addition to GCs, some of the galaxies in our sample may host a few UCDs \citep[see e.g.][]{Hasegan05}, some of which are known to host LMXBs \citep{Dabringhausen12}. If present, most of these compact galaxies will be included in our rather broad GC selection criteria, which includes sources up to luminosities of $L_{z}\sim6.5\times10^{6}L_{\odot}$ \citep[assuming M$_{z,\odot}$=4.51,][]{Sivakoff07}. For this paper, the distinction between UCDs and bright GCs is less important than ensuring that we detect and remove them from our sample of LMXBs in the fields of the galaxies. Given their bright magnitudes, all UCDs that are present will be reliably detected and removed from our analysis. 

\begin{figure}
 \centering
 \includegraphics[width=86mm,angle=270]{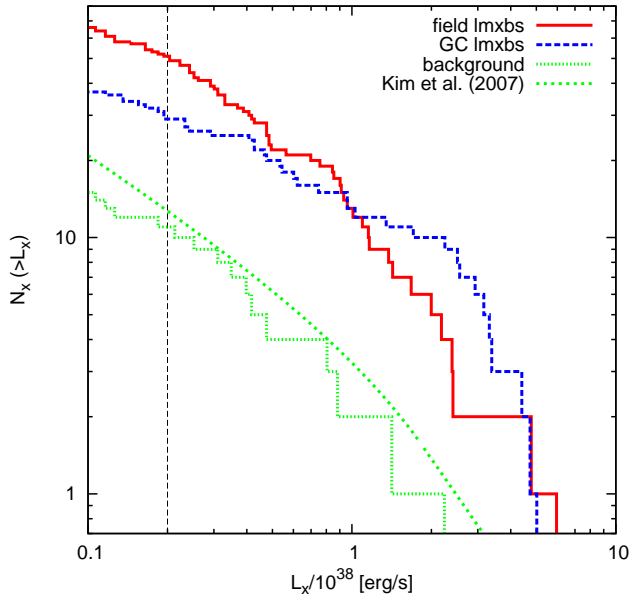} 
 \caption{The cumulative X-ray luminosity function (XLF) of field LMXBs (solid-red line), GC LMXBs (dashed-blue line) and background X-ray sources (dotted-green line) in NGC~4594. The black dashed line indicates the 90\% completeness limit for these X-ray observations. Also plotted is the prediced XLF of background X-ray sources from \citet{Kim07}.}
 \label{fig:ngc4594} 
\end{figure}

Optical counterparts to the X-ray sources in these galaxies were identified in these optical catalogs using their calibrated WCS locations. This was done using the software {\sc topcat/stilts} \citep{Taylor06}. We experimented with different matching radii and found that a matching radius of 0.6$\arcsec$ detected most real counterparts while producing few random false matches. The number of random false matches present was estimated by shifting our source catalog by $\pm$10$\arcsec$ in RA and DEC and averaging the number of matches found. This predicts 3-7 false matches in the galaxy catalogs, corresponding to $\sim5\%$ of the matches being potentially spurious. Using larger matching radii did not increase the number of detected sources by significantly more than predicted from random matches. We split the matched sources in to GC LMXBs and other counterparts (background galaxies or foreground stars). Previous work has already searched for GC counterparts to the X-ray sources in some of the fields covered by our study. We find good agreement with these previous identifications. The other matched sources are assumed to be background AGN and their X-ray luminosity function (XLF) for each galaxy was found to be in good agreement with that observed in other fields by \citet[][using their `broad bandpass' relationship]{Kim07}. At fainter luminosities we identify fewer AGN sources than predicted. However, this corresponded well with the detection limits of these data. We therefore believe that we directly identify (and remove from our later analysis) the majority of GCs and background AGN. 

As an example of the X-ray populations in these galaxies, we show in Figure \ref{fig:ngc4594} the XLF of the field LMXB (solid-red line), GC LMXB (dashed-blue line) and background X-ray source (dotted-green line) populations in NGC~4594. As discussed above, the background population (those X-ray sources with non-GC like optical counterparts) is in good agreement with that expected from the study of \citet{Kim07}. The figure also highlights the well-known result that early-type galaxies typically have similar numbers of GC and field LMXBs \citep[e.g.][]{Angelini01,Kundu07}. 

The X-ray sources that are confirmed to have no optical counterparts represent the population of LMXBs in the field of these galaxies. This is the population we consider in the subsequent analysis. The final numbers of field LMXBs (and GC LMXBs/background sources) that are identified in each galaxy are listed in Table \ref{tab:galaxy_data2}.

\section{Field LMXB populations}
\label{sec:results} 

\subsection{The field LMXB XLF of these galaxies}
\label{sec:xlf}

\begin{figure}
 \centering
 \includegraphics[width=88mm,angle=270]{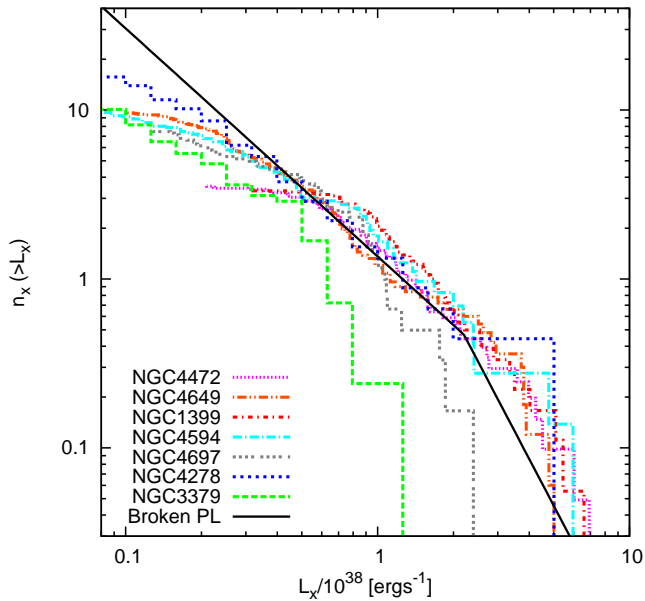} 
 \caption{The XLF of field LMXBs in our sample of galaxies, $n_{x}(>\!L_{x})=N_{x}/L_{K}(>L_{x})$. The number of X-ray sources in each galaxy is scaled by the K-band luminosity covered ($L_{K}$). The black line shows the broken power-law as described by Equation \ref{equ:bpl}, scaled to fit NGC~4278. It can be seen that both the shape of the XLF and normalized number of sources is remarkably similar for most of the galaxies. \\}
 \label{fig:xlf} 
\end{figure}

The matching process discussed in Section \ref{sec:matching} results in a clean population of field LMXBs in each of the galaxies in our sample. Figure \ref{fig:xlf} shows the cumulative X-ray luminosity function for these field LMXBs in seven of the galaxies studied. In this figure, $n_{x}$ is the number of LMXBs (N$_{x}$) scaled by the K-band light (\lk) covered, such that $n_{x}=N_{x}/(\lk\times10^{10}L_{K\odot}$). It can be seen that the flattening at the faint ends of the XLFs is in good agreement with the completeness limits quoted in Table \ref{tab:galaxy_data2}. 

Figure \ref{fig:xlf} shows that the field LMXB XLF is remarkably similar among these galaxies, both in terms of shape and normalized number of sources. The one possibly different galaxy, NGC~3379, is discussed further below. In addition to plotting the stellar-light normalized XLFs, we can also compare them to analytical forms for the XLF. Significant literature is devoted to fitting functional forms to the XLFs of galaxies (including those in our sample) and we do not repeat this previous analysis. Two functions are generally used to represent a galaxy's XLF. The first is a simple power-law of the form: 

\begin{equation}
 \label{equ:pl}
 N(>L_{x}) \propto L_{x}^{-\alpha}
\end{equation}

This single power-law, with an exponent $\alpha$=2.0, has previously been found to represent the bright end of the XLF ($L_{x}>$10$^{38}$) relatively well \citep[e.g.][]{Humphrey08}. However, for these galaxies, where the deep \chandra observations allow the XLF to be studied to fainter luminosities than are often possible, this single power-law poorly represents the observed XLFs. The XLF is found to be much flatter at lower \lx and we find much better agreement with a broken power-law of the form: 

\begin{equation}  \label{equ:bpl}
 N(>L_{x}) \propto \begin{cases}
    (L_{x}/L_{b})^{-\alpha_{1}}, & \text{if $L_{x}>L_{b}$}.\\
    (L_{x}/L_{b})^{-\alpha_{2}}, & \text{otherwise}.
 \end{cases}
\end{equation}
\\
where $L_{b}$ is the break luminosity between the two power-laws. \citet{Humphrey08} have previously fit such a model to a range of early type galaxies (including those studied here) and found that a broken power-law with $L_{b}$=2.2$\times$10$^{38}$\ergs, $\alpha_{1}$=2.84 and $\alpha_{2}$=1.4 provides a good representation of the different galaxy's XLFs. We plot this function, scaled to fit NGC~4278's XLF, in Figure \ref{fig:xlf}. It can be seen that Equation \ref{equ:bpl} represents the high completeness regions of the different galaxies XLFs relatively well. For the two deepest XLF's, those of NGC~3379 and NGC~4278, there are suggestions that the XLF flattens slightly at around $1-2\times10^{37}$\ergs. While a flatter power-law at these faintest luminosities may be genuine, consideration of this is beyond the scope of this paper. This is because we only consider X-ray sources with \lx~$>2\times10^{37}$\ergs in the subsequent analysis. 

Notable from Figure \ref{fig:xlf} is that, at all luminosities, $n_{x}$ is remarkably similar between the different galaxies -- with the possible exception of NGC~3379, which has less sources, particularly at high \lx. The reason for this single outlier remains uncertain. We note that the color of this galaxy is similar to the other galaxies in our sample and it is thought to have a similarly old age, suggesting the low $n_{x}$ is not a stellar population effect. Also, as discussed in Section \ref{sec:imf}, the proposed variations in IMF with galaxy mass, should result in this galaxy having a larger, not smaller, number of LMXBs. 

\begin{figure}
 \centering
 \includegraphics[height=88mm,angle=270]{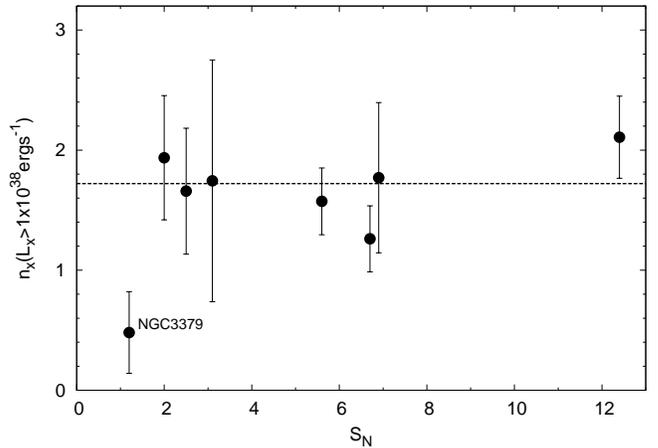} 
 \caption{Normalized number of LMXBs ($n_{x}$) with $L_{x}>10^{38}{\rm ergs^{-1}}$, as a function of globular cluster specific frequency (S$_{N}$). The dashed line represents a constant $n_{x}$ fit to all galaxies except NGC~3379. It can be seen that the data are in good agreement with a fixed $n_{x}$. This suggests that NGC~3379 is an outlier from the other galaxies, rather than following a trend with S$_{N}$.}
 \label{fig:nx_sn} 
\end{figure}

A previously proposed explanation for NGC~3379's low $n_{x}$ is based on the only obvious difference between this and the other galaxies, its relatively low number of GCs. This could result in a lower number of field LMXBs if a significant fraction of these were either ejected from GCs or formed in a GC that was subsequently disrupted. Because GC LMXBs represent a population of LMXBs that is unrelated to the evolution of the field population, it is important to consider whether such a population is significant. \citet{Kim09} suggested a relationship between $n_{x}$ and the GC specific frequency, S$_{N}=N_{\rm GCs}\times10^{0.4(M_{V}+15)}$ based on three galaxies: NGC~3379, NGC~4278 and NGC~4697. In a larger sample of early type galaxies, \citet{Irwin05} found no evidence for such a correlation in their data. Our sample of confirmed field LMXBs allows us to test for such a relationship based on larger samples of galaxies and LMXBs. In Figure \ref{fig:nx_sn}, we plot $n_{x}(L_{x}\!>\!10^{38}{\rm ergs^{-1}})$ as a function of S$_{N}$ for all eight galaxies in our sample. The S$_{N}$ for each galaxy are taken from \citet{Ashman98} for all galaxies except NGC~7457 \citep[which we take from][]{Hargis11}. It can be seen that there is little evidence for a trend between $n_{x}$ and S$_{N}$ and, with the exception of NGC~3379, the data are in excellent agreement with a constant $n_{x}$. Additionally, a number of other observations suggest that the majority of field LMXBs have non GC origins. For example: the LMXB population of the Milky Way is associated with the Galaxy's disk rather than its halo population \citep{Liu01}, indicating that it formed in the field and not in GCs; the radial profile of field LMXBs in a sample of early type galaxies has been observed to trace the I-band light profile better than the GC profile, suggesting a primordial field origin \citep{Kundu07}; also there is evidence of differences in the XLFs of GC and field LMXB populations \citep{Voss09,Zhang11}, suggesting different origins. These indicate that the contamination from non-primordial LMXBs in these galaxies is likely to be negligible.

\subsection{The normalized number of field LMXBs}
\label{sec:trends}

\begin{figure}
 \centering
 \includegraphics[width=88mm,angle=0]{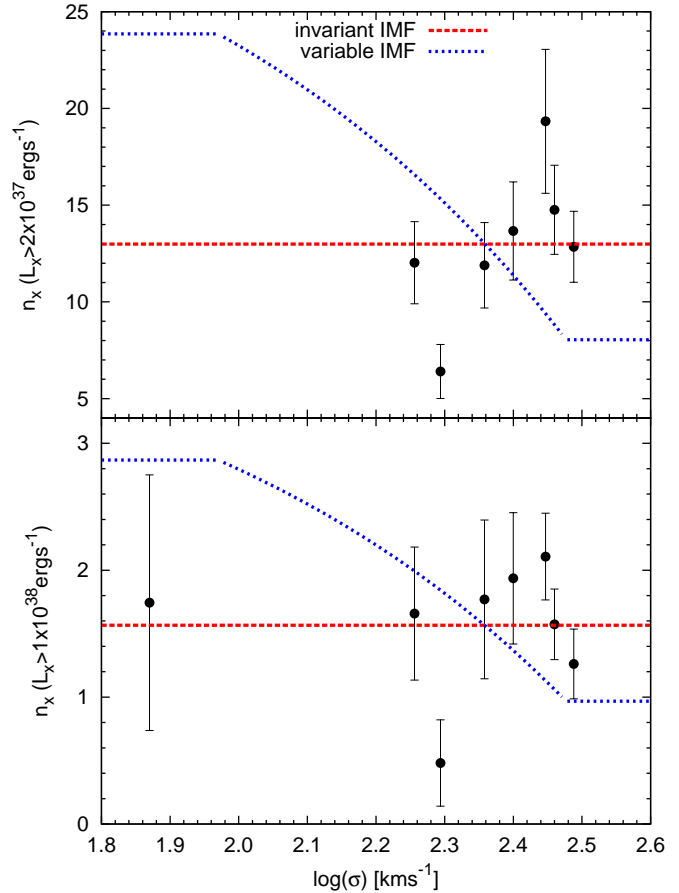} 
 \caption{Total number of field LMXBs in each galaxy (scaled by \lk) with \lx$>2\times10^{37}$\ergs ($n_{x,37}$, top) and \lx$>$10$^{38}$\ergs ($n_{x,38}$, bottom) as a function of the galaxy's velocity dispersion ($\sigma$). The number of sources is scaled by the amount of K-band stellar light covered, such that an invariant IMF among these galaxies predicts a constant $n_{x}$. This case, discussed in Section \ref{sec:constant_imf} is represented by the red dashed line. The blue dotted line shows the predicted variation of $n_{x}$ with $\sigma$ assuming that the IMF varies as required to explain the observed $M/L$ ratio trends, as discussed in Section \ref{sec:variable_imf}. It can be seen that the constant IMF model provides a better representation of the data than the variable IMF model. \\}
 \label{fig:N_M_Lx1e38} 
\end{figure}

\begin{figure}
 \centering
 \includegraphics[width=88mm,angle=0]{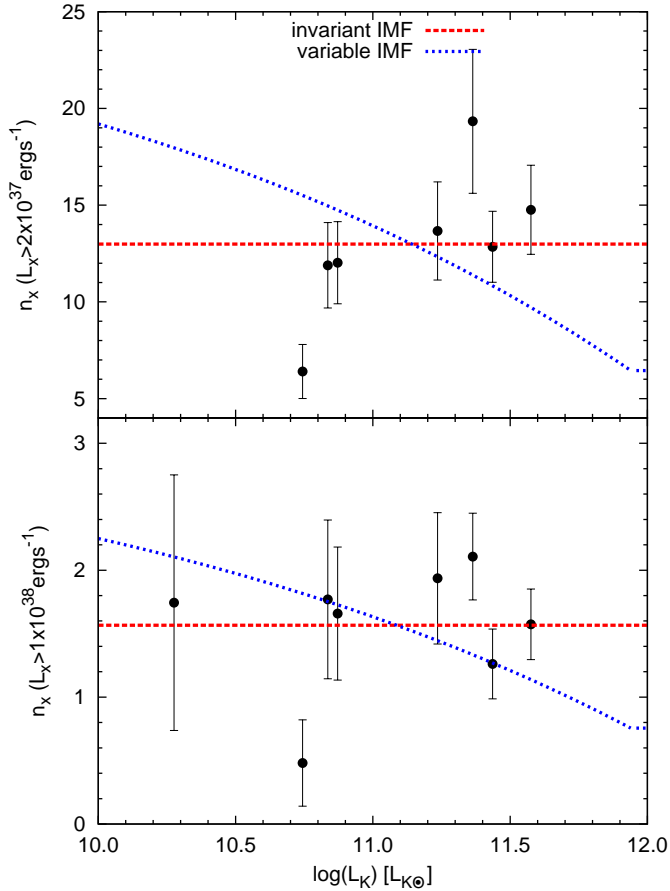} 
 \caption{Similar to Figure \ref{fig:N_M_Lx1e38}, but showing the number of field LMXBs (per \lk) in each galaxy with \lx$>2\times10^{37}$\ergs ($n_{x,37}$, top) and \lx$>$10$^{38}$\ergs ($n_{x,38}$, bottom) as a function of the galaxy's K-band luminosity (\lk). The red-dashed line shows the prediction for an invariant IMF. The blue-dotted line shows the expected variation in $n_{x}$, if the IMF varies as a function of \lk to explain the observed $M/L$ variation (see Section \ref{sec:variable_imf} for details). It can be seen that the observed $n_{x}$ in these galaxies is in better agreement with the constant IMF prediction. \\}
 \label{fig:N_M_Lx2e37} 
\end{figure}

In Figures \ref{fig:N_M_Lx1e38} and \ref{fig:N_M_Lx2e37}, we show how $n_{x}$ varies as a function of a galaxy's velocity dispersion ($\sigma$, measured at 1~kpc by the studies referenced in Table \ref{tab:galaxy_data}) and total K-band luminosity \citep[\lk, from the 2MASS LGA,][]{Jarrett03}. We consider $n_{x}$ down to two different X-ray limits. The bottom panels show the bright X-ray sources in each galaxy, those with $L_{x}>10^{38}$\ergs (hereafter, $n_{x,38}$). The X-ray populations for all of the galaxies should be complete to this limit and therefore require no assumptions about the XLF. However, this limit only considers the brightest end of the XLF and therefore includes only a small fraction of the total LMXB population. This is a particular issue for the lower luminosity galaxies which only have a small number of sources above this limit. In order to take advantage of the deeper detection limits for some of the galaxies, we plot in the top panels of Figures \ref{fig:N_M_Lx1e38} and \ref{fig:N_M_Lx2e37} the number of LMXBs with $L_{x}>2\times10^{37}$\ergs (hereafter, $n_{x,37}$). For NGC~3379 and NGC~4278, the XLFs should be complete to this limit. For the other galaxies we extrapolate their LMXB population by assuming the universal broken power-law XLF discussed in Section \ref{sec:xlf} and fitting it to the galaxies XLF above its detection limit. The LMXB population to this limit therefore has improved statistics for some of the galaxies but, for others, it is more susceptible to systematic errors due to extrapolating the XLF. We do not quote an $n_{x,37}$ for NGC~7457 because its high \lx detection limit and low number of sources do not allow us to accurately extrapolate its XLF to these lower luminosities. The number of LMXBs as a function of $\sigma$ and \lk is found to be in good agreement between the two detection limits. 

Apparent from Figures \ref{fig:N_M_Lx1e38} and \ref{fig:N_M_Lx2e37} is that the normalized number of field LMXBs is similar among these galaxies. We note that this is in agreement with the previous result of \citet[][see e.g.\ their Figure 5]{Kundu03}, who studied a smaller sample of galaxies (NGC~1399, NGC~3115, NGC4365 and NGC~4472). In these figures, we also show the predicted variation in $n_{x}$ for an invariant IMF (red-dashed line) and an IMF which becomes increasingly bottom heavy as a function of galaxy mass (blue-dotted line). It can be seen that the similar $n_{x}$ observed is in better agreement with a fixed IMF than a variable one. In the next section, we discuss these predictions and their consistency with these data. 

\section{LMXB constraints on IMF variations}
\label{sec:imf}

In this section, we discuss the influence of a galaxy's IMF on its LMXB population. Specifically, we predict the variation in $n_{x}$, as a function of galaxy mass, for an invariant IMF (Section \ref{sec:constant_imf}) and an IMF that varies systematically with galaxy mass (Section \ref{sec:variable_imf}). We compare these predictions with the data presented in the previous section and discuss the implications for the inferred IMFs of these galaxies.

\subsection{An invariant IMF}
\label{sec:constant_imf}

If the IMF is invariant among all of these galaxies, the number of black holes and neutron stars should simply scale with the mass of the stellar population. One would therefore predict a constant $n_{x}$ as a function of galaxy mass. Such a scenario is represented in Figures \ref{fig:N_M_Lx1e38} and \ref{fig:N_M_Lx2e37} by the horizontal red dashed line. The formation of LMXBs in the field of a galaxy is still a relatively poorly understood process, likely involving a period of common envelope evolution. Because accurate constraints on the number of LMXBs formed in a stellar population are not available from such theories, we fit the scaling of the line in Figures \ref{fig:N_M_Lx1e38} and \ref{fig:N_M_Lx2e37} to the data. It can be seen that a universal number of LMXBs, per unit K-band luminosity, provides a good representation of the data. Running a $\chi^{2}$ test between $n_{x,38}$ and this model we find $\chi^{2}/\nu$=2.0, which is consistent with the data. For $n_{x,37}$, the smaller error bars suggest a poorer fit with $\chi^{2}/\nu$=4.3. This is inconsistent with the data with a significance of just over 3$\sigma$. 

It is clear from Figures \ref{fig:N_M_Lx1e38} and \ref{fig:N_M_Lx2e37}, that the main outlier from the constant $n_{x}$ model is NGC~3379, which has a relatively low number of LMXBs (as discussed above). Rerunning our tests, but excluding NGC~3379, we find $\chi^{2}/\nu$=0.96 and 0.81 (for $n_{x,38}$ and $n_{x,37}$, respectively) -- confirming that the other seven galaxies are in excellent agreement with the constant IMF model. While the reason for NGC~3379 differing from the other galaxies is still uncertain, it is unlikely related to the proposed IMF variations. This is because its relatively low mass should produce more LMXBs under the proposed IMF variations, not fewer as observed. Additionally, NGC~4278 and NGC~4697 have quite similar masses to NGC~3379, so a mass dependent IMF would be expected to effect all of these galaxies in a similar way. We note that, even including NGC~3379, the data are only inconsistent (by around 3$\sigma$) for $n_{x,37}$. The population of NGC~3379 is better constrained to this fainter limit, thanks to its relative proximity and deep observations. However, it should be noted that (with the exception of NGC~4278) the other galaxies are not complete to this detection limit and so their $n_{x}$s are more susceptible to systematic errors in the scaling. Clearly it will be important to increase this work to larger samples of galaxies and to try to push their XLFs to deeper limits. In the future, larger samples should help to identify whether NGC~3379 is a true outlier or whether significant variations in $n_{x}$ are also present in other galaxies. 

\subsection{A variable IMF - as a function of galaxy mass} 
\label{sec:variable_imf}

In this section, we predict and test the effect of the proposed IMF variations with galaxy luminosity and velocity dispersion on the number of LMXBs in elliptical galaxies. To predict the variation with galaxy mass, we note that it is now thought that the proposed IMF variation must explain the observed \ml variation with galaxy mass, as observed in the fundamental plane \citep{Cappellari12}. Below, we briefly review the fundamental plane of elliptical galaxies and the observed $M/L$ variations with galaxy parameters before calculating the required IMF variation and resulting variation in the number of black holes (BHs) and neutron stars (NSs). 

One of the key features of elliptical galaxies is that they have a very tight relation between their velocity dispersion, effective radius and surface brightness: $R_e \propto \sigma^{1.24} I_e^{-0.82}$, known as the ``fundamental plane'' \citep{Dressler87, Djorgovski87}. It has been known since the discovery of the fundamental plane that the observed scaling between velocity dispersion and the parameters relating to the effective radius and surface brightness for early-type galaxies deviates from that expected from the Virial theorem. This deviation is such that the \ml for early-type galaxies increases systematically with increasing galaxy mass, luminosity, and velocity dispersion. Many studies have quantified this trend \citep[see e.g. the recent comprehensive work by][ and references therein]{Graves10}, finding for example $(M/L)_V \propto \sigma^{0.95}$ \citep{Graves10}, $(M/L)_r \propto \sigma^{0.72}$ \citep{Cappellari13}, $(M/L)_V \propto L_V^{0.23}$ \citep{Mobasher99}, and $(M/L)_K \propto L_K^{0.186}$ \citep[e.g.][]{Pahre98, Mobasher99,LaBarbera10}. Importantly, these trends are all much larger than can be accounted for by the known stellar populations differences among early-type galaxies. Specifically, many studies over the years have investigated the role increasing metallicity with increasing galaxy luminosity. As described in \citet{Graves10}, this falls far short of accounting for the increase in \ml with increasing $\sigma$. They find that stellar population effects account for only about one-quarter of the observed trend, and that the remaining ``tilt'' in the fundamental plane follows \ml $ \propto \sigma^{0.65}$. Additionally, the near-infrared work cited above is mostly immune to metallicity, so \ml $\propto L_K^{0.186}$ also describes the mass-to-light trend with metallicity accounted for. The long-standing question has been what causes this systematic increase in \ml with increasing early-type galaxy mass, luminosity and velocity dispersion. Moreover, because the fundamental plane is very narrow, whatever the cause, it must work very systematically across the range of early-type galaxies. 

Two possible mechanisms are commonly presented to account for the tilt of the fundamental plane. The first, is that higher velocity dispersion early-type galaxies may have systematically larger fractions of dark matter in the inner regions measured at an effective radius or so. The alternative explanation is that the IMF may vary, with higher luminosity and larger velocity dispersion early-type galaxies having a systematically larger \ml ratio because of systematic changes in the IMF. These possibilities are outlined in many papers \citep[e.g.][]{Renzini93, Zepf96, Treu10, Graves10, Cappellari12}. It was recently argued by \citet{Cappellari12} that dark matter variations can not explain the observations. If this is the case, then the trends observed must be explained by an IMF that varies systematically as a function of galaxy mass. 

To investigate how the IMF of a galaxy must vary to explain the above \ml relations, we adopt a model in which the galaxy's IMF is a mixture of a standard Kroupa IMF and a power-law IMF with $x$=2.8. We then vary the ratio of these components to explain the observed trends of \ml with velocity dispersion and luminosity. Using the stellar population synthesis models of \citet[][rerun for a power-law IMF with x=2.8]{Maraston05}, for a 10~Gyr simple stellar population with solar or half solar metallicity, we find that the difference in the K-band \ml resulting from the different IMFs is:
\\
\begin{equation}
R_{M/L}=\frac{(M/\lk)_{2.8}}{(M/\lk)_{{\rm kro}}}=2.3. 
\end{equation}
\\
This ratio is the same as that used by \citet{Cappellari12} and is quite insensitive to the exact age and metallicity of the stellar population. We proceed by first considering the number of NSs and BHs that are produced from the evolution of the two IMFs considered. In order to normalize the two IMFs to the same total mass, we consider the Kroupa IMF of the form: 

\begin{equation}
 \frac{dM}{dm} = N_{0,{\rm kro}} \begin{cases}
    m \times m^{-2.3}, & \text{$m>0.5M_{\odot}$}\\
    m \times m^{-1.3}, & \text{$0.1M_{\odot}<m<0.5M_{\odot}$}\\
 \end{cases}
\end{equation}
\\
and the power-law IMF with exponent $x$=2.8: 
\begin{equation}
 \frac{dM}{dm} = N_{0,2.8} \times m \times m^{-2.8}
\end{equation}
\\
The two normalization constants $N_{0}$ can then be found for each IMF by integrating over the total stellar mass range (0.1-100$M_{\odot}$) and normalizing to a total mass of 1$M_{\odot}$. Performing this integration yields $N_{0,{\rm kro}}=0.225$ and $N_{0,2.8}=0.127$. We can now find the fraction of the total number of stars in these two IMFs that evolve into NSs and BHs by assuming these form from stars with initial masses $>$8\Msun: 

\begin{equation} \label{equ:M_kro}
 N_{\rm kro}(M>\!8\Msun) = 0.225\int_{8}^{100}m^{-2.3}dm = 0.01116
\end{equation}

\begin{equation} \label{equ:M_2.8}
 N_{2.8}(M>8\Msun) = 0.127\int_{8}^{100}m^{-2.8}dm = 0.00165
\end{equation}
\\ 
The ratio of \ref{equ:M_kro} and \ref{equ:M_2.8} will give the difference in the number of NSs and BHs present in a stellar population with the same total mass but formed from a Kroupa rather than an $x$=2.8 IMF. However, we require this ratio for a constant luminosity population, because we normalize our data by this easier to observe parameter. We therefore find that the ratio of BHs and NSs in the Kroupa to the $x$=2.8 IMF is:

\begin{equation}\label{equ:R_NSBH}
R_{\rm NS/BH} = \frac{0.01116}{0.00165}\times \frac{1}{R_{M/L}} = 2.9 
\end{equation}

Next we calculate how the ratio of the Kroupa IMF component to an $x$=2.8 IMF component has to vary, as a function of $\sigma$ to explain the observed variation in the $M/L$ ratio, $M/L \propto \sigma^{0.72}$ \citep{Cappellari13}. For galaxies with low masses, those with $\sigma\sim90$kms$^{-1}$, their observed $M/L$ requires that their IMF must be similar to a Kroupa IMF. For galaxies with higher $\sigma$, we then increase the $x$=2.8 component so as to match the observed increase in their $M/L$ ratios. The resulting fraction of the IMF composed of the $x$=2.8 component, $F_{2.8}$, as a function of $\sigma$ is given by: 

\begin{align}\label{equ:F_kro}
\begin{split}
 F_{2.8}(\sigma)  &=   \frac{1}{(R_{M/L}-1)} \left( \frac{(M/L)_{\sigma}}{(M/L)_{\rm kro}}-1 \right) \\
                           &=   \frac{1}{(R_{M/L}-1)} \left( \frac{\sigma^{0.72}}{95^{0.72}}-1 \right) 
\end{split}
\end{align}
\\
In this way $F_{2.8}$=0.0 and the IMF is purely Kroupa, when $\sigma$=95~kms$^{-1}$ (as defined) and $F_{2.8}$=1.0 and the IMF is purely an $x$=2.8 IMF, when $\sigma$=300~kms$^{-1}$. As an additional test, we also consider the the variation of $F_{2.8}$ required to explain the observed relationship between the $M/L$ and ${\rm L_{K}}$, \ml $\propto L_{K}^{0.186}$ \citep{LaBarbera10}. We again note that galaxies at the low mass end, those with luminosities of \lk$\sim$1$\times10^{10}L_{K\odot}$ are expected to have Kroupa like IMFs. Following a similar process used for the $\sigma$ relationship, the fraction of the $x$=2.8 IMF as a function of \lk is found to be:  

\begin{align}\label{equ:F_28}
\begin{split}
 F_{2.8}(L_{\rm K})  &=   \frac{1}{(R_{M/L}-1)} \left( \frac{L_{\rm K}^{0.186}}{(1\times10^{10})^{0.186}}-1 \right) 
\end{split}
\end{align}

Finally, having found how the ratio of the IMF components varies as a function of $\sigma$ and \lk, we use equation \ref{equ:R_NSBH} to calculate the resulting variation in the number of NSs and BHs: 

\begin{align}
\begin{split}
\label{equ:nx}
\frac{N_{\rm NS/BH}(\sigma /L_{\rm K})}{N_{\rm NS/BH}({\rm kro})} 
&= 
\frac{F_{2.8}}{R_{\rm NS/BH}} + F_{\rm kro} \\
&= 
1-\left( 1-\frac{1}{R_{\rm NS/BH}} \right) F_{2.8}
\end{split}
\end{align}
\\
Where we have noted that the fraction of the Kroupa IMF,  $F_{\rm kro}=1-F_{2.8}$. This relationship, combined with equations \ref{equ:F_kro} and \ref{equ:F_28}, is used to produce the predicted numbers of LMXBs as a function of $\sigma$ and \lk in Figures \ref{fig:N_M_Lx1e38} and \ref{fig:N_M_Lx2e37} (the blue dotted lines). 

It can be seen from Figures \ref{fig:N_M_Lx1e38} and \ref{fig:N_M_Lx2e37} that the data are in poor agreement with these predicted trends. To test these predictions against the observed number of LMXBs, we run $\chi^{2}$ tests between the data and the predicted trends with both $\sigma$ and \lk. The reduced $\chi^{2}$ statistics obtained for the $\sigma$ and \lk relations to the $n_{x,37}$ data are $\chi^{2}/\nu$=10.5 and $\chi^{2}/\nu$=9.5, respectively. For the 5 degrees of freedom, the variable IMF models are therefore strongly rejected, with a confidence of 6.8$\sigma$ and 6.4$\sigma$. Considering the data for only the brightest sources in each galaxy, $n_{x,38}$, we find that, for the six degrees of freedom, $\chi^{2}/\nu$=4.5 for $\sigma$ and $\chi^{2}/\nu$=3.5 for \lk. This is again significantly inconsistent, although the confidence is lower due to the larger observational uncertainties. Furthermore, we note that even if we exclude the galaxy NGC~3379 from our fits (which is an outlier from both the invariant and variable IMF models), the $n_{x,37}$ data are still in much poorer agreement with the variable IMF than the invariant one, with $\chi^{2}/\nu$=4.8 for $\sigma$ and $\chi^{2}/\nu$=2.9 for \lk (c.f. $\chi^{2}/\nu$=0.8 for $\sigma$ and $\chi^{2}/\nu$=0.96 for \lk for an invariant IMF, see Section \ref{sec:constant_imf}). 

\subsection{Other effects on the LMXB population}
\label{sec:other}

Our goal is to test for IMF variations in early-type galaxies as a function of their $\sigma$ and \lk by comparing their normalized number of field LMXBs. To do this, we need to consider the potential effect of other properties that vary with $\sigma$ and \lk in these galaxies. One candidate is metallicity ($Z$), which is well-known to increase with increasing $\sigma$ and \lk. Fortunately, we can calculate the effect of this on the number of LMXBs. \citet{Thomas10} found that this trend in early-type galaxies is such that [Z/H]=-1.34+0.65$log(\sigma)$. Thus over the range covered by our sample of galaxies (with detection limits, $L_{X} > 10^{37}$\ergs), the expected increase in [Z/H] is only 0.14~dex. Although we do not know the dependence of field LMXBs on metallicity directly, we do know the dependence within GCs, where the number of LMXBs scales as $Z^{0.32}$ \citep[][]{Smits06,Sivakoff07}. Thus, the expectation is that the increasing metallicity of brighter, higher velocity dispersion galaxies will increase the LMXB numbers by only about $10\%$. This falls far short of the proposed increase due to changing the IMF (which is $\sim300\%$ over this range of $\sigma$). Thus, the observed metallicity dependence of elliptical galaxies does not effect our analysis.

Another property of elliptical galaxies to consider is age. The galaxies in our sample were often selected for
X-ray observations because stellar populations studies indicated that they had a uniformly old age \citep[e.g.][]{Kim09}. Therefore, age effects would seem to be an unlikely source of variation. However, in a statistical sense, lower velocity dispersion galaxies are found to have slightly younger ages \citep[e.g.][]{Graves07,Thomas10}. Adopting the \citet{Thomas10} relation between galaxy age and $\sigma$, gives an increase in age from about 9 to 11 Gyr as $\sigma$ increases from 180 to 290 kms$^{-1}$ \citep[similar ages are also predicted from the relation of][]{Graves07}. We note that there are few constraints on the variation of LMXB numbers over this range of ages. Therefore, an extremely steep age dependence of LMXB formation at these old ages can not be completely excluded. However, it seems difficult to achieve a large change over such a small range of ages. In particular, to hide the proposed variation in the number of LMXBs due to IMF variations, any such change would have to be dramatic (a factor 3 or so from around 9 to 11~Gyrs) and carefully tuned to decrease with $\sigma$ to produce the constant LMXB number observed.

Therefore, the comparison of the normalized number of field LMXBs in different early-type galaxies is a direct test of the ratio of the number of (now evolved) massive stars to the number of approximately solar mass stars that are now dominating the light in these old galaxies. Our results presented here demonstrate that this ratio appears to be constant for the most part across a wide range of early-type galaxy masses. The simplest variations of the IMF, in which its slope varies with galaxy mass, would therefore appear to be ruled out. It is important to emphasize though, that the LMXB test presented here does not directly test the ratio of the slightly less than solar mass stars currently dominating the light of early-type galaxies to very low mass stars that contribute almost no light. Thus it is possible to construct an IMF variation in which the IMF is invariant for nearly all masses, but with a varying contribution from very low mass ($<$0.3$M_{\odot}$) stars. Such a population of very low-mass stars only would have to have the same spatial distribution as the ``normal'' IMF population in order to satisfy the constraints from dynamics and strong lensing that mass follows light. Whether such an IMF is physically plausible remains to be seen. However, if one wishes to preserve IMF variations to explain previous work in the context of this study, a solution of this kind is required. Alternatively, the IMF may not vary, and other astrophysical explanations of near-infrared line features in early-type galaxies may be found. An invariant IMF would also require new a different explanation for the dynamical observations and some lensing results, such as returning to dark matter arguments.

\section{Conclusions}

In this paper, we use the number of field LMXBs per stellar luminosity to investigate the ratio of the number of high mass stars ($\gtrsim$8\Msun) to low mass stars ($<$1\Msun) that were formed in a sample of local early-type galaxies.  We find that the XLFs and normalized number of field LMXBs ($n_{x}$) are remarkably similar among the galaxies observed. 

We consider the implications of this result for the IMF, specifically predicting the expectations from an invariant IMF and an IMF which becomes increasingly bottom heavy as a galaxy's mass increases. The latter variation is motivated as an explanation for the correlation observed between the \ml of an early-type galaxy and its $\sigma$ and \lk. We find that the data are more consistent with an invariant IMF than a variable one. Indeed, we show that the data are inconsistent with a situation where galaxy IMFs change from a Kroupa IMF for low mass galaxies to an IMF which is the sum of a Kroupa plus an $x$=2.8 power-law for higher mass galaxies. We conclude that there is no evidence in the LMXB populations of these galaxies for the ratio of high mass stars increasing with decreasing galaxy mass. Such a correlation would have been expected under the previously proposed IMF variations that were invoked to explain the observed spectra and dynamics of these galaxies. 

We note that one galaxy in our sample of eight, NGC~3379, is also inconsistent with a fixed IMF. While we can not identify a reason for the relatively low number of sources in this galaxy, we note that systematic IMF variations with mass can not explain it. This galaxy highlights the need to extend this work to larger samples of galaxies. This will be possible with new and deeper \chandra observations of more galaxies. Additionally, new \hst mosaics of nearby early-type galaxies that have \chandra data will allow us to study larger regions of the galaxies. Particularly important will be new \chandra data for galaxies in the low mass range, where IMF effects should be largest and produce relatively large numbers of LMXBs. Additionally, deeper data for NGC~7457 will allow us to more accurately constrain its LMXB population.

\section*{Acknowledgments}

We thank the anonymous referee of this paper for careful consideration and providing detailed comments that were beneficial to the final version. We thank Tana Joseph for providing us with a copy of her catalog of X-ray sources in NGC~4472, Nicola Brassington for providing us with an original copy of her catalog of X-ray sources in NGC~3379 and Jay Strader for helpful discussions related to this paper. We also thank Charlie Conroy, Pieter van Dokkum, Ignacio Ferreras, Pavel Kroupa and Russell Smith for helpful comments on the arXiv e-print version of this paper. 

MBP and SEZ acknowledge support from NASA through the ADAP grant NNX11AG12G and through the Chandra award AR4-15007X. AK acknowledges support for this work provided by NASA through Chandra awards GO0-11111A and AR1-12009X.

This research has made use of NASA's Astrophysics Data System.

\bibliographystyle{apj_w_etal}
\bibliography{bibliography_etal}

\begin{thebibliography}{88}
\expandafter\ifx\csname natexlab\endcsname\relax\def\natexlab#1{#1}\fi

\bibitem[{{Angelini} {et~al.}(2001){Angelini}, {Loewenstein}, \&
  {Mushotzky}}]{Angelini01}
{Angelini}, L., {Loewenstein}, M., \& {Mushotzky}, R.~F. 2001, ApJ, 557, L35

\bibitem[{{Ashman} \& {Zepf}(1998)}]{Ashman98}
{Ashman}, K.~M. \& {Zepf}, S.~E. 1998, {Globular Cluster Systems}, Cambridge
  astrophysics series ; 30 (Cambridge University Press, Cambridge,~UK)

\bibitem[{{Bastian} {et~al.}(2010){Bastian}, {Covey}, \& {Meyer}}]{Bastian10}
{Bastian}, N., {Covey}, K.~R., \& {Meyer}, M.~R. 2010, \araa, 48, 339

\bibitem[{{Bertin} \& {Arnouts}(1996)}]{Bertin96}
{Bertin}, E. \& {Arnouts}, S. 1996, A\&AS, 117, 393

\bibitem[{{Blakeslee} {et~al.}(2009){Blakeslee}, {Jord{\'a}n}, {Mei},
  {C{\^o}t{\'e}}, {Ferrarese}, {Infante}, {Peng}, {Tonry}, \&
  {West}}]{Blakeslee09}
{Blakeslee}, J.~P. {et~al.} 2009, \apj, 694, 556

\bibitem[{{Blakeslee} {et~al.}(2001){Blakeslee}, {Lucey}, {Barris}, {Hudson},
  \& {Tonry}}]{Blakeslee01}
{Blakeslee}, J.~P., {Lucey}, J.~R., {Barris}, B.~J., {Hudson}, M.~J., \&
  {Tonry}, J.~L. 2001, \mnras, 327, 1004

\bibitem[{{Brassington} {et~al.}(2008){Brassington}, {Fabbiano}, {Kim},
  {Zezas}, {Zepf}, {Kundu}, {Angelini}, {Davies}, {Gallagher}, {Kalogera},
  {Fragos}, {King}, {Pellegrini}, \& {Trinchieri}}]{Brassington08}
{Brassington}, N.~J. {et~al.} 2008, \apjs, 179, 142

\bibitem[{{Brassington} {et~al.}(2009){Brassington}, {Fabbiano}, {Kim},
  {Zezas}, {Zepf}, {Kundu}, {Angelini}, {Davies}, {Gallagher}, {Kalogera},
  {Fragos}, {King}, {Pellegrini}, \& {Trinchieri}}]{Brassington09}
---. 2009, \apjs, 181, 605

\bibitem[{{Cappellari} {et~al.}(2012){Cappellari}, {McDermid}, {Alatalo},
  {Blitz}, {Bois}, {Bournaud}, {Bureau}, {Crocker}, {Davies}, {Davis}, {de
  Zeeuw}, {Duc}, {Emsellem}, {Khochfar}, {Krajnovi{\'c}}, {Kuntschner},
  {Lablanche}, {Morganti}, {Naab}, {Oosterloo}, {Sarzi}, {Scott}, {Serra},
  {Weijmans}, \& {Young}}]{Cappellari12}
{Cappellari}, M. {et~al.} 2012, \nat, 484, 485

\bibitem[{{Cappellari} {et~al.}(2013){Cappellari}, {McDermid}, {Alatalo},
  {Blitz}, {Bois}, {Bournaud}, {Bureau}, {Crocker}, {Davies}, {Davis}, {de
  Zeeuw}, {Duc}, {Emsellem}, {Khochfar}, {Krajnovi{\'c}}, {Kuntschner},
  {Morganti}, {Naab}, {Oosterloo}, {Sarzi}, {Scott}, {Serra}, {Weijmans}, \&
  {Young}}]{Cappellari13}
---. 2013, \mnras, 432, 1862

\bibitem[{{Carter} {et~al.}(1986){Carter}, {Visvanathan}, \&
  {Pickles}}]{Carter86}
{Carter}, D., {Visvanathan}, N., \& {Pickles}, A.~J. 1986, \apj, 311, 637

\bibitem[{{Cenarro} {et~al.}(2003){Cenarro}, {Gorgas}, {Vazdekis}, {Cardiel},
  \& {Peletier}}]{Cenarro03}
{Cenarro}, A.~J., {Gorgas}, J., {Vazdekis}, A., {Cardiel}, N., \& {Peletier},
  R.~F. 2003, \mnras, 339, L12

\bibitem[{{Chabrier}(2003)}]{Chabrier03}
{Chabrier}, G. 2003, \pasp, 115, 763

\bibitem[{{Clark}(1975)}]{Clark75}
{Clark}, G.~W. 1975, ApJ, 199, L143

\bibitem[{{Cohen}(1978)}]{Cohen78}
{Cohen}, J.~G. 1978, \apj, 221, 788

\bibitem[{{Conroy} \& {van Dokkum}(2012)}]{Conroy12b}
{Conroy}, C. \& {van Dokkum}, P.~G. 2012, \apj, 760, 71

\bibitem[{{Couture} \& {Hardy}(1993)}]{Couture93}
{Couture}, J. \& {Hardy}, E. 1993, \apj, 406, 142

\bibitem[{{Dabringhausen} {et~al.}(2012){Dabringhausen}, {Kroupa},
  {Pflamm-Altenburg}, \& {Mieske}}]{Dabringhausen12}
{Dabringhausen}, J., {Kroupa}, P., {Pflamm-Altenburg}, J., \& {Mieske}, S.
  2012, \apj, 747, 72

\bibitem[{{de Vaucouleurs} {et~al.}(1991){de Vaucouleurs}, {de Vaucouleurs},
  {Corwin}, {Buta}, {Paturel}, \& {Fouqu{\'e}}}]{deVaucouleurs91}
{de Vaucouleurs}, G., {de Vaucouleurs}, A., {Corwin}, Jr., H.~G., {Buta},
  R.~J., {Paturel}, G., \& {Fouqu{\'e}}, P. 1991, {Third Reference Catalogue of
  Bright Galaxies.} (Springer, New York, USA)

\bibitem[{{Djorgovski} \& {Davis}(1987)}]{Djorgovski87}
{Djorgovski}, S. \& {Davis}, M. 1987, \apj, 313, 59

\bibitem[{{Dressler} {et~al.}(1987){Dressler}, {Lynden-Bell}, {Burstein},
  {Davies}, {Faber}, {Terlevich}, \& {Wegner}}]{Dressler87}
{Dressler}, A., {Lynden-Bell}, D., {Burstein}, D., {Davies}, R.~L., {Faber},
  S.~M., {Terlevich}, R., \& {Wegner}, G. 1987, \apj, 313, 42

\bibitem[{{Faber} \& {French}(1980)}]{Faber80}
{Faber}, S.~M. \& {French}, H.~B. 1980, \apj, 235, 405

\bibitem[{{Fabian} {et~al.}(1975){Fabian}, {Pringle}, \& {Rees}}]{Fabian75}
{Fabian}, A.~C., {Pringle}, J.~E., \& {Rees}, M.~J. 1975, MNRAS, 172, 15P

\bibitem[{{Ferreras} {et~al.}(2013){Ferreras}, {La Barbera}, {de la Rosa},
  {Vazdekis}, {de Carvalho}, {Falc{\'o}n-Barroso}, \&
  {Ricciardelli}}]{Ferreras13}
{Ferreras}, I., {La Barbera}, F., {de la Rosa}, I.~G., {Vazdekis}, A., {de
  Carvalho}, R.~R., {Falc{\'o}n-Barroso}, J., \& {Ricciardelli}, E. 2013,
  \mnras, 429, L15

\bibitem[{{Gebhardt} {et~al.}(2007){Gebhardt}, {Lauer}, {Pinkney}, {Bender},
  {Richstone}, {Aller}, {Bower}, {Dressler}, {Faber}, {Filippenko}, {Green},
  {Ho}, {Kormendy}, {Siopis}, \& {Tremaine}}]{Gebhardt07}
{Gebhardt}, K. {et~al.} 2007, \apj, 671, 1321

\bibitem[{{Geha} {et~al.}(2013){Geha}, {Brown}, {Tumlinson}, {Kalirai},
  {Simon}, {Kirby}, {VandenBerg}, {Mu{\~n}oz}, {Avila}, {Guhathakurta}, \&
  {Ferguson}}]{Geha13}
{Geha}, M. {et~al.} 2013, \apj, 771, 29

\bibitem[{{Gonzaga} {et~al.}(2013)}]{Gonzaga13}
{Gonzaga}, S. {et~al.} 2013, {ACS Data Handbook: Version 7.0} (Baltimore:
  STScI)

\bibitem[{{Graves} \& {Faber}(2010)}]{Graves10}
{Graves}, G.~J. \& {Faber}, S.~M. 2010, \apj, 717, 803

\bibitem[{{Graves} {et~al.}(2007){Graves}, {Faber}, {Schiavon}, \&
  {Yan}}]{Graves07}
{Graves}, G.~J., {Faber}, S.~M., {Schiavon}, R.~P., \& {Yan}, R. 2007, \apj,
  671, 243

\bibitem[{{G{\"u}ltekin} {et~al.}(2012){G{\"u}ltekin}, {Cackett}, {Miller}, {Di
  Matteo}, {Markoff}, \& {Richstone}}]{Gultekin12}
{G{\"u}ltekin}, K., {Cackett}, E.~M., {Miller}, J.~M., {Di Matteo}, T.,
  {Markoff}, S., \& {Richstone}, D.~O. 2012, \apj, 749, 129

\bibitem[{{Ha{\c s}egan} {et~al.}(2005){Ha{\c s}egan}, {Jord{\'a}n},
  {C{\^o}t{\'e}}, {Djorgovski}, {McLaughlin}, {Blakeslee}, {Mei}, {West},
  {Peng}, {Ferrarese}, {Milosavljevi{\'c}}, {Tonry}, \& {Merritt}}]{Hasegan05}
{Ha{\c s}egan}, M. {et~al.} 2005, \apj, 627, 203

\bibitem[{{Hargis} {et~al.}(2011){Hargis}, {Rhode}, {Strader}, \&
  {Brodie}}]{Hargis11}
{Hargis}, J.~R., {Rhode}, K.~L., {Strader}, J., \& {Brodie}, J.~P. 2011, \apj,
  738, 113

\bibitem[{{Humphrey} \& {Buote}(2008)}]{Humphrey08}
{Humphrey}, P.~J. \& {Buote}, D.~A. 2008, \apj, 689, 983

\bibitem[{{Irwin}(2005)}]{Irwin05}
{Irwin}, J.~A. 2005, \apj, 631, 511

\bibitem[{{Jardel} {et~al.}(2011){Jardel}, {Gebhardt}, {Shen}, {Fisher},
  {Kormendy}, {Kinzler}, {Lauer}, {Richstone}, \& {G{\"u}ltekin}}]{Jardel11}
{Jardel}, J.~R. {et~al.} 2011, \apj, 739, 21

\bibitem[{{Jarrett} {et~al.}(2003){Jarrett}, {Chester}, {Cutri}, {Schneider},
  \& {Huchra}}]{Jarrett03}
{Jarrett}, T.~H., {Chester}, T., {Cutri}, R., {Schneider}, S.~E., \& {Huchra},
  J.~P. 2003, \aj, 125, 525

\bibitem[{{Jensen} {et~al.}(2003){Jensen}, {Tonry}, {Barris}, {Thompson},
  {Liu}, {Rieke}, {Ajhar}, \& {Blakeslee}}]{Jensen03}
{Jensen}, J.~B. {et~al.} 2003, \apj, 583, 712

\bibitem[{{Jord{\'a}n} {et~al.}(2004)}]{Jordan04}
{Jord{\'a}n}, A. {et~al.} 2004, ApJ, 613, 279

\bibitem[{{Jord{\'a}n} {et~al.}(2007)}]{Jordan07}
---. 2007, ApJ, 671, L117

\bibitem[{Joseph(2013)}]{Joseph13}
Joseph, T. 2013, PhD thesis, University of Southampton

\bibitem[{{Kim} {et~al.}(2009){Kim}, {Fabbiano}, {Brassington}, {Fragos},
  {Kalogera}, {Zezas}, {Jord{\'a}n}, {Sivakoff}, {Kundu}, {Zepf}, {Angelini},
  {Davies}, {Gallagher}, {Juett}, {King}, {Pellegrini}, {Sarazin}, \&
  {Trinchieri}}]{Kim09}
{Kim}, D.-W. {et~al.} 2009, \apj, 703, 829

\bibitem[{{Kim} {et~al.}(2006){Kim}, {Kim}, {Fabbiano}, {Lee}, {Park},
  {Geisler}, \& {Dirsch}}]{Kim06}
{Kim}, E., {Kim}, D., {Fabbiano}, G., {Lee}, M.~G., {Park}, H.~S., {Geisler},
  D., \& {Dirsch}, B. 2006, ApJ, 647, 276

\bibitem[{{Kim} {et~al.}(2007){Kim}, {Lee}, {Geisler}, {Sarajedini}, {Park},
  {Hwang}, {Harris}, {Seguel}, \& {von Hippel}}]{Kim07}
{Kim}, S.~C. {et~al.} 2007, AJ, 134, 706

\bibitem[{{Kroupa}(2001)}]{Kroupa01}
{Kroupa}, P. 2001, \mnras, 322, 231

\bibitem[{{Kundu} {et~al.}(2002){Kundu}, {Maccarone}, \& {Zepf}}]{Kundu02}
{Kundu}, A., {Maccarone}, T.~J., \& {Zepf}, S.~E. 2002, ApJ, 574, L5

\bibitem[{{Kundu} {et~al.}(2007){Kundu}, {Maccarone}, \& {Zepf}}]{Kundu07}
---. 2007, ApJ, 662, 525

\bibitem[{{Kundu} {et~al.}(2003){Kundu}, {Maccarone}, {Zepf}, \&
  {Puzia}}]{Kundu03}
{Kundu}, A., {Maccarone}, T.~J., {Zepf}, S.~E., \& {Puzia}, T.~H. 2003, ApJ,
  589, L81

\bibitem[{{La Barbera} {et~al.}(2010){La Barbera}, {de Carvalho}, {de La Rosa},
  \& {Lopes}}]{LaBarbera10}
{La Barbera}, F., {de Carvalho}, R.~R., {de La Rosa}, I.~G., \& {Lopes},
  P.~A.~A. 2010, \mnras, 408, 1335

\bibitem[{{La Barbera} {et~al.}(2013){La Barbera}, {Ferreras}, {Vazdekis}, {de
  la Rosa}, {de Carvalho}, {Trevisan}, {Falc{\'o}n-Barroso}, \&
  {Ricciardelli}}]{LaBarbera13}
{La Barbera}, F. {et~al.} 2013, \mnras, 433, 3017

\bibitem[{{Li} {et~al.}(2010){Li}, {Spitler}, {Jones}, {Forman}, {Kraft}, {Di
  Stefano}, {Tang}, {Wang}, {Gilfanov}, \& {Revnivtsev}}]{Li10}
{Li}, Z. {et~al.} 2010, \apj, 721, 1368

\bibitem[{{Liu}(2011)}]{Liu11}
{Liu}, J. 2011, \apjs, 192, 10

\bibitem[{{Liu} {et~al.}(2001){Liu}, {van Paradijs}, \& {van den
  Heuvel}}]{Liu01}
{Liu}, Q.~Z., {van Paradijs}, J., \& {van den Heuvel}, E.~P.~J. 2001, A\&A,
  368, 1021

\bibitem[{{Luo} {et~al.}(2013){Luo}, {Fabbiano}, {Strader}, {Kim}, {Brodie},
  {Fragos}, {Gallagher}, {King}, \& {Zezas}}]{Luo13}
{Luo}, B. {et~al.} 2013, \apjs, 204, 14

\bibitem[{{Maccarone} {et~al.}(2003){Maccarone}, {Kundu}, \&
  {Zepf}}]{Maccarone03}
{Maccarone}, T.~J., {Kundu}, A., \& {Zepf}, S.~E. 2003, \apj, 586, 814

\bibitem[{{Maraston}(2005)}]{Maraston05}
{Maraston}, C. 2005, MNRAS, 362, 799

\bibitem[{{Mieske} {et~al.}(2008){Mieske}, {Hilker}, {Jord{\'a}n}, {Infante},
  {Kissler-Patig}, {Rejkuba}, {Richtler}, {C{\^o}t{\'e}}, {Baumgardt}, {West},
  {Ferrarese}, \& {Peng}}]{Mieske08a}
{Mieske}, S. {et~al.} 2008, \aap, 487, 921

\bibitem[{{Mieske} \& {Kroupa}(2008)}]{Mieske08b}
{Mieske}, S. \& {Kroupa}, P. 2008, \apj, 677, 276

\bibitem[{{Mobasher} {et~al.}(1999){Mobasher}, {Guzman}, {Aragon-Salamanca}, \&
  {Zepf}}]{Mobasher99}
{Mobasher}, B., {Guzman}, R., {Aragon-Salamanca}, A., \& {Zepf}, S. 1999,
  \mnras, 304, 225

\bibitem[{{Pahre} {et~al.}(1998){Pahre}, {Djorgovski}, \& {de
  Carvalho}}]{Pahre98}
{Pahre}, M.~A., {Djorgovski}, S.~G., \& {de Carvalho}, R.~R. 1998, \aj, 116,
  1591

\bibitem[{{Paolillo} {et~al.}(2011){Paolillo}, {Puzia}, {Goudfrooij}, {Zepf},
  {Maccarone}, {Kundu}, {Fabbiano}, \& {Angelini}}]{Paolillo11}
{Paolillo}, M. {et~al.} 2011, \apj, 736, 90

\bibitem[{{Peacock} {et~al.}(2010){Peacock}, {Maccarone}, {Kundu}, \&
  {Zepf}}]{Peacock10b}
{Peacock}, M.~B., {Maccarone}, T.~J., {Kundu}, A., \& {Zepf}, S.~E. 2010,
  \mnras, 407, 2611

\bibitem[{{Renzini} \& {Ciotti}(1993)}]{Renzini93}
{Renzini}, A. \& {Ciotti}, L. 1993, \apjl, 416, L49

\bibitem[{{Saglia} {et~al.}(2000){Saglia}, {Kronawitter}, {Gerhard}, \&
  {Bender}}]{Saglia00}
{Saglia}, R.~P., {Kronawitter}, A., {Gerhard}, O., \& {Bender}, R. 2000, \aj,
  119, 153

\bibitem[{{Saglia} {et~al.}(2002){Saglia}, {Maraston}, {Thomas}, {Bender}, \&
  {Colless}}]{Saglia02}
{Saglia}, R.~P., {Maraston}, C., {Thomas}, D., {Bender}, R., \& {Colless}, M.
  2002, \apjl, 579, L13

\bibitem[{{Salpeter}(1955)}]{Salpeter55}
{Salpeter}, E.~E. 1955, \apj, 121, 161

\bibitem[{{S{\'a}nchez-Bl{\'a}zquez} {et~al.}(2006){S{\'a}nchez-Bl{\'a}zquez},
  {Gorgas}, {Cardiel}, \& {Gonz{\'a}lez}}]{Sanchez-Blazquez06b}
{S{\'a}nchez-Bl{\'a}zquez}, P., {Gorgas}, J., {Cardiel}, N., \& {Gonz{\'a}lez},
  J.~J. 2006, \aap, 457, 809

\bibitem[{{Schlegel} {et~al.}(1998){Schlegel}, {Finkbeiner}, \&
  {Davis}}]{Schlegel98}
{Schlegel}, D.~J., {Finkbeiner}, D.~P., \& {Davis}, M. 1998, ApJ, 500, 525

\bibitem[{{Sil'chenko}(2006)}]{Silchenko06}
{Sil'chenko}, O.~K. 2006, \apj, 641, 229

\bibitem[{{Sirianni} {et~al.}(2005){Sirianni}, {Jee}, {Ben{\'{\i}}tez},
  {Blakeslee}, {Martel}, {Meurer}, {Clampin}, {De Marchi}, {Ford}, {Gilliland},
  {Hartig}, {Illingworth}, {Mack}, \& {McCann}}]{Sirianni05}
{Sirianni}, M. {et~al.} 2005, \pasp, 117, 1049

\bibitem[{{Sivakoff} {et~al.}(2008){Sivakoff}, {Jord{\'a}n}, {Juett},
  {Sarazin}, \& {Irwin}}]{Sivakoff08}
{Sivakoff}, G.~R., {Jord{\'a}n}, A., {Juett}, A.~M., {Sarazin}, C.~L., \&
  {Irwin}, J.~A. 2008, ArXiv e-prints

\bibitem[{{Sivakoff} {et~al.}(2007){Sivakoff}, {Jord{\'a}n}, {Sarazin},
  {Blakeslee}, {C{\^o}t{\'e}}, {Ferrarese}, {Juett}, {Mei}, \&
  {Peng}}]{Sivakoff07}
{Sivakoff}, G.~R. {et~al.} 2007, \apj, 660, 1246

\bibitem[{{Smith} \& {Lucey}(2013)}]{Smith13}
{Smith}, R.~J. \& {Lucey}, J.~R. 2013, \mnras, 434, 1964

\bibitem[{{Smits} {et~al.}(2006){Smits}, {Maccarone}, {Kundu}, \&
  {Zepf}}]{Smits06}
{Smits}, M., {Maccarone}, T.~J., {Kundu}, A., \& {Zepf}, S.~E. 2006, A\&A, 458,
  477

\bibitem[{{Taylor}(2006)}]{Taylor06}
{Taylor}, M.~B. 2006, in Astronomical Society of the Pacific Conference Series,
  Vol. 351, Astronomical Data Analysis Software and Systems XV, ed.
  C.~{Gabriel}, C.~{Arviset}, D.~{Ponz}, \& S.~{Enrique}, 666

\bibitem[{{Terlevich} \& {Forbes}(2002)}]{Terlevich02}
{Terlevich}, A.~I. \& {Forbes}, D.~A. 2002, \mnras, 330, 547

\bibitem[{{Thomas} {et~al.}(2010){Thomas}, {Maraston}, {Schawinski}, {Sarzi},
  \& {Silk}}]{Thomas10}
{Thomas}, D., {Maraston}, C., {Schawinski}, K., {Sarzi}, M., \& {Silk}, J.
  2010, \mnras, 404, 1775

\bibitem[{{Tonry} {et~al.}(2001){Tonry}, {Dressler}, {Blakeslee}, {Ajhar},
  {Fletcher}, {Luppino}, {Metzger}, \& {Moore}}]{Tonry01}
{Tonry}, J.~L. {et~al.} 2001, \apj, 546, 681

\bibitem[{{Trager} {et~al.}(2000){Trager}, {Faber}, {Worthey}, \&
  {Gonz{\'a}lez}}]{Trager00}
{Trager}, S.~C., {Faber}, S.~M., {Worthey}, G., \& {Gonz{\'a}lez}, J.~J. 2000,
  \aj, 120, 165

\bibitem[{{Treu} {et~al.}(2010){Treu}, {Auger}, {Koopmans}, {Gavazzi},
  {Marshall}, \& {Bolton}}]{Treu10}
{Treu}, T., {Auger}, M.~W., {Koopmans}, L.~V.~E., {Gavazzi}, R., {Marshall},
  P.~J., \& {Bolton}, A.~S. 2010, \apj, 709, 1195

\bibitem[{{van Dokkum} \& {Conroy}(2010)}]{vanDokkum10}
{van Dokkum}, P.~G. \& {Conroy}, C. 2010, \nat, 468, 940

\bibitem[{{van Dokkum} \& {Conroy}(2011)}]{vanDokkum11}
---. 2011, \apjl, 735, L13

\bibitem[{{Verbunt} \& {Hut}(1987)}]{Verbunt87}
{Verbunt}, F. \& {Hut}, P. 1987, in IAU Symposium, Vol. 125, The Origin and
  Evolution of Neutron Stars, ed. {D.~J.~Helfand \& J.-H.~Huang}, 187--+

\bibitem[{{Villegas} {et~al.}(2010){Villegas}, {Jord{\'a}n}, {Peng},
  {Blakeslee}, {C{\^o}t{\'e}}, {Ferrarese}, {Kissler-Patig}, {Mei}, {Infante},
  {Tonry}, \& {West}}]{Villegas10}
{Villegas}, D. {et~al.} 2010, \apj, 717, 603

\bibitem[{{Voss} \& {Gilfanov}(2007)}]{Voss07}
{Voss}, R. \& {Gilfanov}, M. 2007, \aap, 468, 49

\bibitem[{{Voss} {et~al.}(2009){Voss}, {Gilfanov}, {Sivakoff}, {Kraft},
  {Jord{\'a}n}, {Raychaudhury}, {Birkinshaw}, {Brassington}, {Croston},
  {Evans}, {Forman}, {Hardcastle}, {Harris}, {Jones}, {Juett}, {Murray},
  {Sarazin}, {Woodley}, \& {Worrall}}]{Voss09}
{Voss}, R. {et~al.} 2009, \apj, 701, 471

\bibitem[{{Weidner} {et~al.}(2013){Weidner}, {Ferreras}, {Vazdekis}, \& {La
  Barbera}}]{Weidner13}
{Weidner}, C., {Ferreras}, I., {Vazdekis}, A., \& {La Barbera}, F. 2013,
  \mnras, 435, 2274

\bibitem[{{Zepf} \& {Silk}(1996)}]{Zepf96}
{Zepf}, S.~E. \& {Silk}, J. 1996, \apj, 466, 114

\bibitem[{{Zhang} {et~al.}(2011){Zhang}, {Gilfanov}, {Voss}, {Sivakoff},
  {Kraft}, {Brassington}, {Kundu}, {Jord{\'a}n}, \& {Sarazin}}]{Zhang11}
{Zhang}, Z. {et~al.} 2011, \aap, 533, A33

\end{thebibliography}

\label{lastpage}

\end{document}